\documentclass[reprint, 
groupedaddress,
superscriptaddress,
longbibliography,
amsmath,amssymb,
aps,
pra,
]{revtex4-2}
\usepackage[colorlinks = true,]{hyperref}
\usepackage{float}
\usepackage{graphicx}
\usepackage{dcolumn}
\usepackage{bm}
\usepackage{xcolor}
\usepackage{braket}
\usepackage{soul}
\usepackage{comment}
\usepackage{amsmath}

\begin{document}
\title{Robust Universal Photon Blockade in a Bimodal Jaynes–Cummings Model via Kerr Nonlinearity}
\author{Chang Guohao}%
\affiliation{School of Physics and Electronic Engineering, Xinjiang Normal University, Urumqi, Xinjiang 830054, China}
\author{Hunduz Halemjan}%
\affiliation{School of Physics and Electronic Engineering, Xinjiang Normal University, Urumqi, Xinjiang 830054, China}
\author{Liu Shangyun}%
\affiliation{School of Physics and Electronic Engineering, Xinjiang Normal University, Urumqi, Xinjiang 830054, China}
\author{Raziya Anwar}%
\affiliation{School of Physics and Electronic Engineering, Xinjiang Normal University, Urumqi, Xinjiang 830054, China}
\author{Ahmad Abliz}%
\email{aahmad@126.com}
\affiliation{School of Physics and Electronic Engineering, Xinjiang Normal University, Urumqi, Xinjiang 830054, China}
\affiliation{Lanzhou Center for Theoretical Physics, Key Laboratory of Theoretical Physics of Gansu Province, Key Laboratory of Quantum Theory and Applications of MoE, Gansu Provincial Research Center for Basic Disciplines of Quantum Physics, Lanzhou University, Lanzhou 730000, China}

\begin{abstract} 
Universal photon blockade in a two-mode Jaynes-Cummings model incorporating third-order Kerr nonlinearity is demonstrated with a single two-level atom coupled to a waveguide microcavity. Realization of this universal photon blockade is attributed to the cooperative effects of field-atom coupling and Kerr nonlinearity. More importantly, this antibunching is found to be robust against the atomic spontaneous emission, driving field strength, and defect-induced cavity mode coupling. The strong antibunching effect in this resonance-driven scheme is essentially different from those without Kerr nonlinearity. Moreover, this work expands the platform for achieving universal photon blockade and reveals the cooperative advantages of nonlinearities in enhancing the purity and brightness of single-photon sources, representing a novel strategy toward high-performance single-photon sources in integrated quantum optical devices.
\end{abstract}

\date{\today}

\maketitle

\section{Introduction}
Operating in a deterministic fashion, single-photon sources enable the triggered generation of light, exclusively yielding one photon per emission cycle. They serve as a key to unlocking and controlling the quantum world, and have potential application in integrated quantum optical devices, e.g., as quantum interferometers \cite{1}, single-photon routers \cite{2,3}, single-photon transistors \cite{4,5}, and non-classical isolators \cite{6,7}. Photon blockade (PB) has attracted extensive research attention in recent decades as an effective method for preparing single-photon sources. PB involves the absorption of a single photon while blocking the absorption of subsequent photons, thereby enabling an ordered successive output of individual photons \cite{8,9,10}. Nonlinear interactions are found to be crucial for achieving PB. PB can be divided into two types—conventional photon blockade (CPB) and unconventional photon blockade (UPB), depending on the underlying mechanism: the former relies on strong nonlinearity-induced anharmonic splitting of energy levels, effectively suppressing multiphoton transitions through spectral inhomogeneity \cite{11,12,13,14,15,16,17}, while the latter exploits destructive quantum interference between multiple paths to inhibit two-photon excitation even under weak coupling, significantly reducing the requirement for strong nonlinearity and enabling the generation of single photons with higher purity \cite{17,18,19,20,21}. Both mechanisms were experimentally verified in cavity quantum electrodynamics (QED) systems coupled to single two-level systems, e.g., traditional QED systems with coupled atoms \cite{12,22}, solid-state quantum systems \cite{13,23,24,25,26,27}, and microwave systems \cite{28,29,30,31}, providing a vital basis for developing novel quantum optical devices.

Recent studies have indicated that the mechanisms of CPB and UPB are not completely independent, implying that the cooperative effects of spectral anharmonicity and quantum interference can yield enhanced PB within specific parameter ranges. Strong nonlinearity-induced energy level splitting provides energy selection for quantum interference pathways, while the destructive interference between these pathways further suppresses residual multiphoton probabilities, thereby overcoming the limitations of single-mechanism approaches. For example, Tang et al. \cite{32} achieved a strong PB in a three-level atom-cavity system by leveraging the synergistic effects of the optical Stark shift and quantum interference. Zhu et al.\cite{33} found a synergistically enhanced PB through spectral anharmonicity and quantum interference which can be tuned via interatomic dipole-dipole interactions in a two-atom Jaynes–Cummings (JC) model, while similar findings were reported by Huang et al. \cite{34} in the same year. Zhu et al. \cite{35} realized PB in a system of two cavities with Kerr nonlinearities by separately controlling the drive ratio and the Kerr nonlinearity ratio between the two cavities. Qiao et al. \cite{36} introduced Kerr nonlinearity into a three-wave mixing system, enabling the coexistence of spectral anharmonicity and quantum interference mechanisms within the strongly coupled region. Zhou et al. \cite{37} recently proposed the concept of “universal photon blockade (universal PB)” in a two-photon JC model by ingeniously combining single-photon resonance with quantum interference using only one nonlinear interaction (cavity-atom coupling) to derive a unified optimization condition. This pioneering work theoretically demonstrated that high-quality photon antibunching can be achieved across the entire parameter range, from strong to moderate to weak nonlinearity, thus opening up new avenues toward high-performance single-photon sources. Universal PB has been demonstrated in several second-order nonlinear systems \cite{38,39}. However, most of the works on the realization of universal PB based on the cooperative effects of two different mechanisms are heavily dependent on specific parameter regimes, which are challenged by stringent experimental requirements \cite{32,33,34,35,36,40,41}. Therefore, the importance of both robust and less stringent universal PB which may facilitate the study of quantum networks and high-quality single photon sources has come to light.

In this paper, we aim to achieve robust universal PB via strong third-order Kerr nonlinearity under the cavity-atom-drive frequency resonant conditions in a bimodal JC model. Specifically, in the weak-coupling regime, the Kerr nonlinearity-induced anharmonicity further suppresses the two-photon population, yielding strong antibunching through the coexistence and interplay of CPB and UPB. The second-order correlation function decreased by more than one order of magnitude while maintaining the same single-photon brightness. In the strong-coupling regime, the resonance-driven energy level anharmonicity further intensifies, yet the system's second-order correlation function remains stable, which is at odds with the CPB. This is due to the fact that the response of the two-level atom pair to the coherent light field becomes saturated under strong coupling, rendering the coupling strength no longer the limiting factor for light field statistics \cite{23,42,43}. The mechanism behind the PB in this work fundamentally differs from those in previous studies in which photon antibunching occurs under resonance-driven and weakly coupled conditions without Kerr nonlinearity \cite{44,45}. Furthermore, we demonstrated that the parameter range for UPB expands with the gradual increase in Kerr nonlinearity, while the blockade effect simultaneously intensifies. More importantly, our universal PB is robust to atomic spontaneous emission, inter-mode coupling strength, and driving field intensity, signiﬁcantly relaxing the stringency of experimental requirements. This study not only expands the physical implementation platform for universal PB but also reveals the advantages of collective interplay of two different mechanisms enhanced by Kerr nonlinearity. The improved purity and brightness of single-photon sources in this scheme may open possibilities for high-performance single-photon manipulation in integrated quantum optical devices.

\section{PHYSICAL MODEL}
\label{sec: PHYSICAL MODEL}
\begin{figure}[t]
    \centering
    \includegraphics[width=1.0\columnwidth]{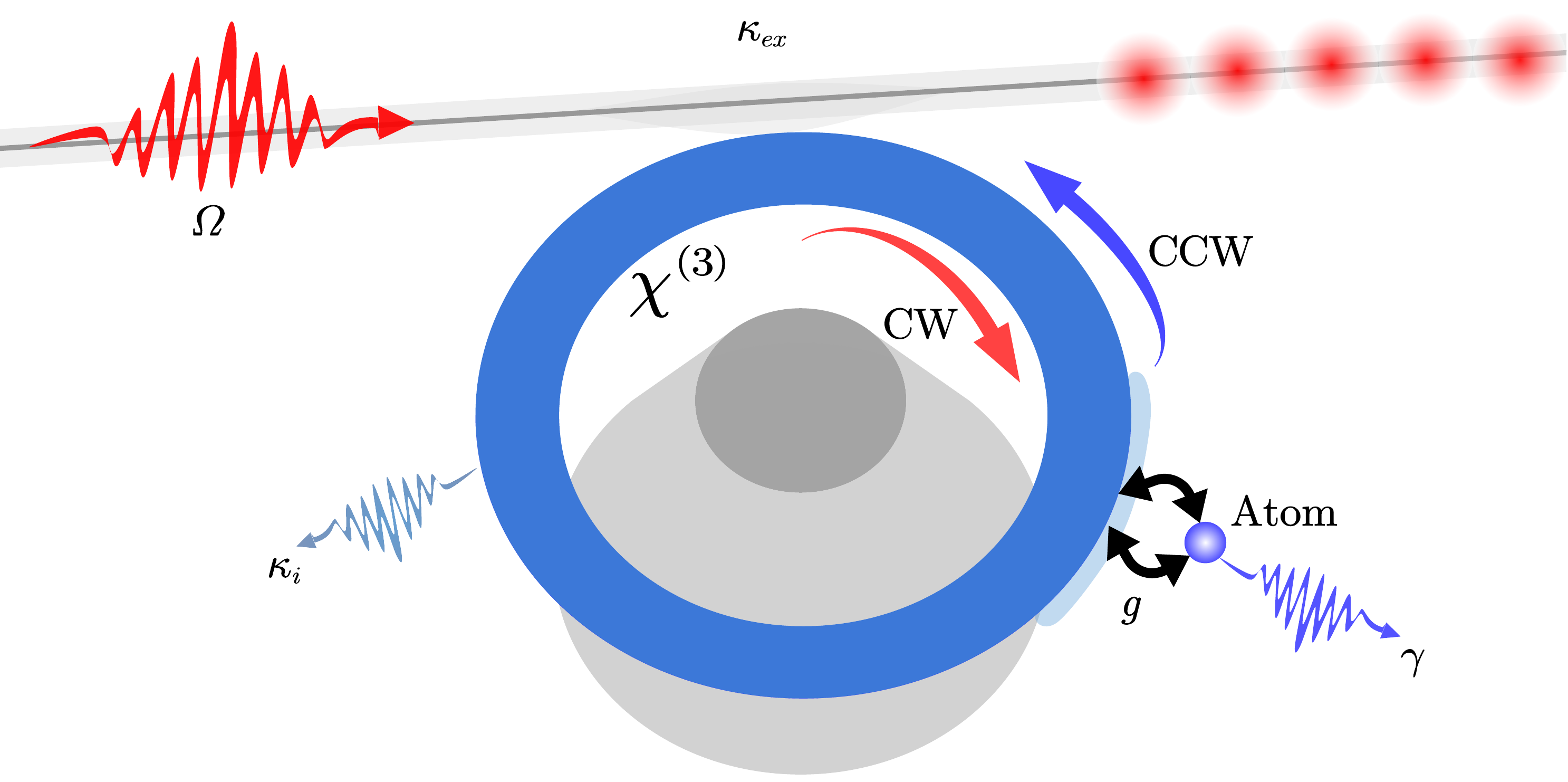}
    \caption{Schematic of a microcavity coupled to a single two-level atomic system. The cavity's CW mode is driven by a tapered fiber waveguide, where $\kappa_{ex}$ is the coupling rate between the waveguide and the microcavity. The WGM microcavity supports two optical modes: a clockwise (CW) propagation mode and a counterclockwise (CCW) propagation mode (indicated by red and blue arrows, respectively), each exhibiting   Kerr nonlinearity. Both modes within the cavity are simultaneously coupled to a single two-level atom with the coupling strength g. The intrinsic decay rate of the microcavity is $\kappa_{i}$, and the spontaneous emission rate of the atom is $\gamma$}.
    \label{fig:system-model}
\end{figure}

As schematically shown in Fig. \ref{fig:system-model}, we study a system comprises a third-order Kerr nonlinearity whispering-gallery-mode (WGM) microcavity with a two-level atom embedded within its effective mode volume. The microcavity supports two degenerate optical modes that correspond to clockwise (CW) and counterclockwise (CCW) propagation directions, both coupled to the atom. The system’s Hamiltonian can be written as follows($\hbar=1$): 

\begin{equation}
\begin{aligned}
\hat{H}_{0} = &\ \omega(\hat{a}_1^\dagger\hat{a}_1+\hat{a}_2^\dagger\hat{a}_2) + \omega_\mathrm{e}\hat{\sigma}_+\hat{\sigma}_- \\
&+ \chi(\hat{a}_1^\dagger\hat{a}_1^\dagger\hat{a}_1\hat{a}_1 + \hat{a}_2^\dagger\hat{a}_2^\dagger\hat{a}_2\hat{a}_2) \\
&+ g(\hat{a}_1^\dagger\hat{\sigma}_- + \hat{a}_1\hat{\sigma}_+) + g(\hat{a}_2^\dagger\hat{\sigma}_- + \hat{a}_2\hat{\sigma}_+) \\
&+ \Omega(\hat{a}_1^\dagger e^{-i\omega_pt} + \hat{a}_1 e^{i\omega_pt}),
\end{aligned}
\end{equation}

where $\hat{a}_{1}$ and $\hat{a}_{2}$ ($\hat{a}_1^{\dagger}$ and $\hat{a}_2^{\dagger}$) represent the annihilation (creation) operators for the two optical modes with frequency $\omega$. $\hat{\sigma}_+$ and $\hat{\sigma}_-$ are atomic transition operator, $\omega_{e}$ is the atomic transition frequency. $\chi$ represents the Kerr nonlinearity; $g$ denotes the interaction strength between atoms and each cavity mode; the final term represents the driving of the CW mode by a coherent light field with the intensity $\Omega$ and the frequency $\omega_{p}$. The cavity-atom resonance, i.e., $\omega=\omega_e$, is assumed. In a rotating frame with the driving field frequency $\omega_p$ (let the detuning between the driving field and cavity mode be $\Delta=\omega-\omega_p$), using the unitary transformation $V=\exp\left[-i\omega_pt(\hat{a}_1^\dagger\hat{a}_1+\hat{a}_2^\dagger\hat{a}_2+\hat{\sigma}_+\hat{\sigma}_-)\right]$, the Hamiltonian can be rewritten as follows:

\begin{equation}
\label{ME_FQ}
\begin{aligned}
\hat{H} ={} & \Delta(\hat{a}_1^\dagger\hat{a}_1+\hat{a}_2^\dagger\hat{a}_2+\hat{\sigma}_+\hat{\sigma}_-) \\
& + \chi(\hat{a}_1^\dagger\hat{a}_1^\dagger\hat{a}_1\hat{a}_1+\hat{a}_2^\dagger\hat{a}_2^\dagger\hat{a}_2\hat{a}_2) \\
& + g(\hat{a}_1^\dagger\hat{\sigma}_-+\hat{a}_1\hat{\sigma}_+) + g(\hat{a}_2^\dagger\hat{\sigma}_-+\hat{a}_2\hat{\sigma}_+) \\
& + \Omega(\hat{a}_1^\dagger+\hat{a}_1).
\end{aligned}
\end{equation}

The system’s dynamical evolution can be achieved by numerically solving the Lindblad master equation:

\begin{equation}
\label{ME_FQ1}
\frac{\partial \hat{\rho}(t)}{\partial t} = -i\left[\hat{H},\hat{\rho}\right] + \kappa D\left[\hat{a}_1\right] + \kappa D\left[\hat{a}_2\right] + \gamma D\left[\hat{\sigma}_-\right].
\end{equation}

Here $\mathcal{D}\begin{bmatrix}\hat{o}\end{bmatrix}$ denotes the Lindblad operator, expressed as $\mathcal{D}\left[\hat{o}\right]=2\hat{o}\hat{\rho}\hat{o}-\hat{o}\hat{o}\hat{\rho}-\hat{\rho}\hat{o}\hat{o}$, where $\hat{o}=\hat{a}_1,\hat{a}_2,\hat{\sigma}_-$, and $\hat{\rho}$ is the system’s reduced density operator. Assuming the total loss rates of CW and CCW modes are equal to $\kappa$, where $\kappa=\kappa_1+\kappa_{ex}=\kappa_2+\kappa_{ex}$, and $\kappa_{1}$ and $\kappa_{2}$ represent the loss rates for the two modes within the cavity, respectively, $\kappa_{ex}$ denotes the coupling loss rate between the waveguide and the microcavity, and $\gamma$ indicates the spontaneous emission rate of the atom. The system’s reduced density operator $\hat{\rho}$ can be obtained by numerically solving Eq. (\ref{ME_FQ}) using Qutip \cite{46,47}. 

As the next step, we focus on the statistical properties of the cavity mode in the WGM via the normalized equal-time second-order correlation function in the steady state $\hat{\rho}_{ss}=\hat{\rho}(t\to\infty)$, which can be obtained by solving the equation for $\partial\hat{\rho}(t)/\partial t=0$. The equal-time second-order correlation function for the CW mode of the system is defined as following: 

\begin{equation}
\label{g2}
g^{(2)}(0)=\frac{\langle\hat{a}_1^{\dagger2}\hat{a}_1^2\rangle}{\langle\hat{a}_1^{\dagger}\hat{a}_1\rangle^2}=\frac{\mathrm{Tr}\left(\rho_{ss}\hat{a}_1^{\dagger2}\hat{a}_1^2\right)}{\left[\mathrm{Tr}(\rho_{ss}\hat{a}_1^{\dagger}\hat{a}_1)\right]^2},
\end{equation}
where $g^{(2)}(0)$ represents the probability of simultaneously observing two photons and can be measured using the Hanbury Brown-Twiss experiment \cite{48}. $g^{(2)}(0)<1$ ($g^{(2)}(0)>1$) is referred to as the sub-Poissonian (super-Poissonian) statistics indicating that the photon exhibits antibunching (bunching), with the ideal photon blockade corresponding to the limit $g^{(2)}(0)\to0$, meaning the high purity of single photons.

\section{PHOTON BLOCKADE}
\label{sec:PHOTON BLOCKADE}
\subsection{Conventional photon blockade}
\label{Conventional photon blockade}

\begin{figure}[t]
    \centering
    \includegraphics[width=1.0\columnwidth]{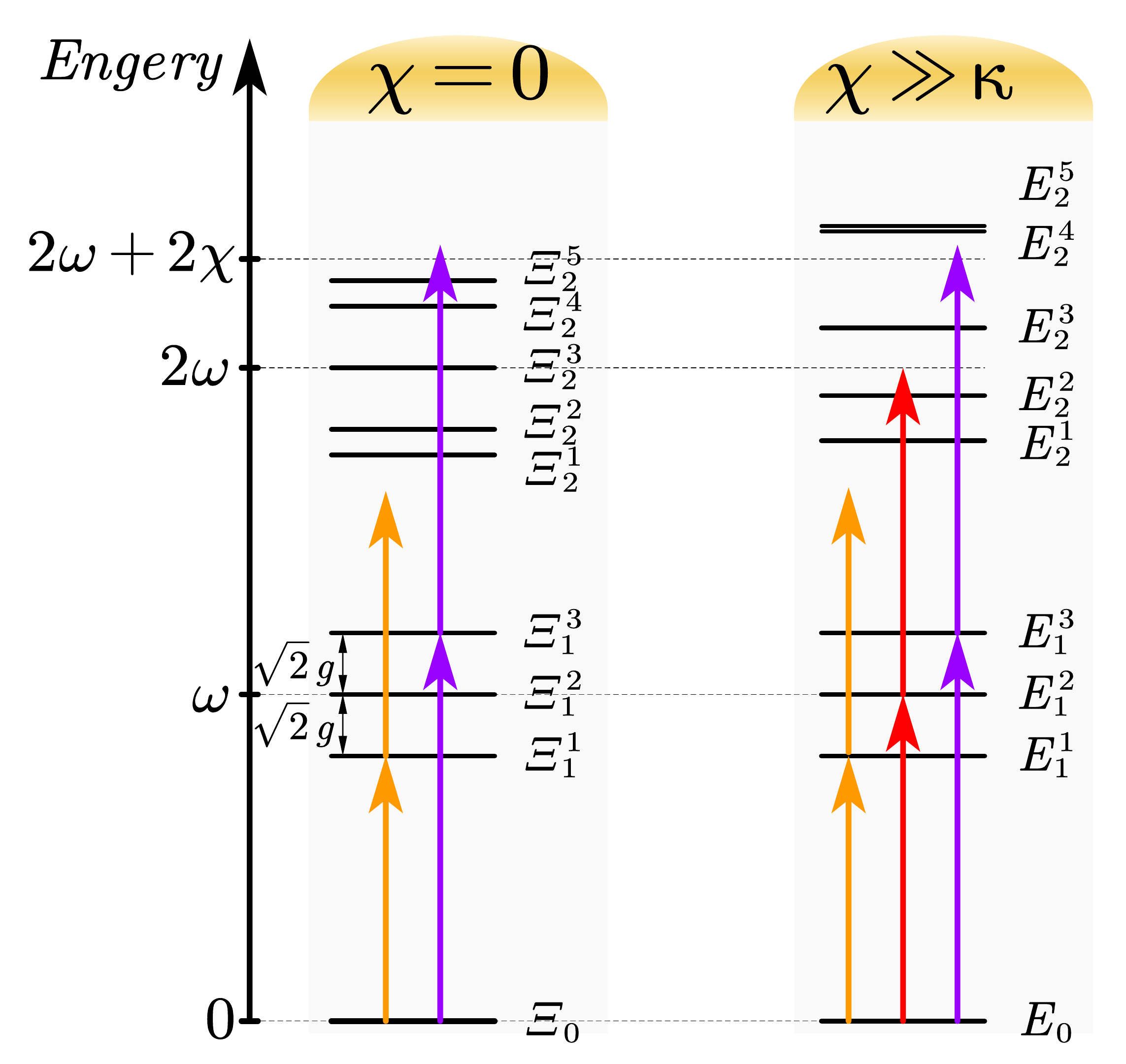}
    \caption{Comparison of the dressed state energy levels for systems with no Kerr nonlinearity (left) and strong Kerr nonlinearity (right) under strong coupling. Three energy levels exist under single excitation, while five levels exist under double excitation.}
    \label{fig:energy-levels}
\end{figure}

CPB is generally explained by the anharmonicity in the intrinsic energy spectrum. CPB without Kerr nonlinearity ($\chi=0$) was studied in \cite{14}. Under strong coupling, CPB occurs only when the detuning satisfies $\Delta=\pm\sqrt{2}g$ (Fig. \ref{fig:energy-levels}). In the case of resonant driving ($\Delta=0$), two-photon resonance ($\Xi_0\to\Xi_2^3$) induces photon-induced tunneling, such that there is no antibunching. However, when strong Kerr nonlinearity ($\chi\gg\kappa$) is present in the system, two-photon excitation energy levels shift, while the single-photon excitation energy levels remain unaffected, leading to CPB even when the driving field is on resonance with the cavity.

\begin{figure}[t]
        \centering
        \includegraphics[width=1.0\linewidth]{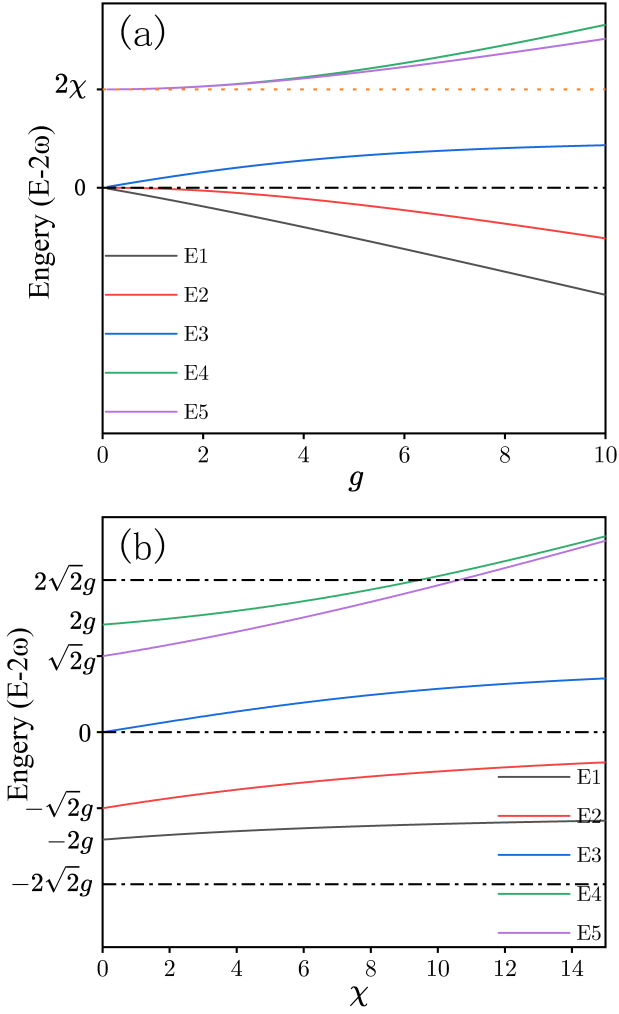}
        \caption{Schematic diagram of the dressed-state energy level structure of the system in the two-photon excitation subspace as functions of the cavity-atom coupling strength $g$ and the Kerr nonlinearity $\chi$. The black dot-dashed line marks the position of the two-excitation energy level on resonance with the single-excitation frequency, while the orange dashed line indicates the energy level at $2(\omega+\chi)$. (a) Under resonant driving, the anharmonicity of the energy levels at zero detuning becomes more pronounced as the cavity-atom coupling strengthens ($\chi=8$). (b) When the cavity-atom coupling is strong, the anharmonicity of the energy levels gradually increases with the enhancement of the Kerr nonlinearity ($g=10$)}

        \label{fig:excitation}
\end{figure}

The anharmonicities in energy levels with the atom-field coupling g and Kerr nonlinearity $\chi$, respectively, are shown in Fig. \ref{fig:excitation}. The eigenspectrum of the dressed state of the system can be obtained from the matrix form of the system Hamiltonian in the two-photon subspace, which is written as:  

\begin{equation}
\hat{H}=\begin{pmatrix}
2\omega+2\chi & 0 & 0 & g\sqrt{2} & 0 \\
0 & 2\omega & 0 & g & g \\
0 & 0 & 2\omega+2\chi & 0 & g\sqrt{2} \\
g\sqrt{2} & g & 0 & 2\omega & 0 \\
0 & g & g\sqrt{2} & 0 & 2\omega
\end{pmatrix}.
\end{equation}

From Fig. \ref{fig:excitation} one can indeed fully confirm that CPB occurs at the resonance ($\Delta=0$) under strong coupling.

Furthermore, it is indicated that by enhancing Kerr nonlinearity or coupling strength one can achieve effective energy level anharmonicity, yielding stronger CPB. The optimal detuning for CPB in this case is:

\begin{equation}
\Delta=0,\quad\pm\sqrt{2}g.
\label{eq:optimal_detuning}
\end{equation}

\subsection{Unconventional photon blockade}

To clarify the specific conditions for destructive interference between the different transition pathways that can enhance the UPB effect, we employ the probability amplitude method for analytically calculating the equal-time second-order functions. Detailed comparisons are made with numerical simulations. According to the quantum trajectory theory, the non-Hermitian effective Hamiltonian of the system can be expressed as follows:

\begin{equation}
H_{\mathrm{eff}}=H-i\frac{\kappa}{2}\hat{a}_1^\dagger\hat{a}_1-i\frac{\kappa}{2}\hat{a}_2^\dagger\hat{a}_2-i\frac{\gamma}{2}\hat{\sigma}_+\hat{\sigma}_-.
\end{equation}

Under weak driving conditions, i.e., $\Omega\ll\kappa$, the system can only excite a small number of photons. The total number of excitations in this case is restricted to $N\leq2$, with $N=\langle\hat{a}_1^\dagger\hat{a}_1+\hat{a}_2^\dagger\hat{a}_2+\hat{\sigma}_+\hat{\sigma}_-\rangle$. Then the wave function describing the system can be approximately written as follows:

\begin{equation}
\label{psi_state}
\begin{aligned}
\mid\psi(t)\rangle = & C_{00g}\mid0,0,g\rangle + C_{10g}\mid1,0,g\rangle + C_{01g}\mid0,1,g\rangle \\
& + C_{00e}\mid0,0,e\rangle + C_{20g}\mid2,0,g\rangle + C_{02g}\mid0,2,g\rangle \\
& + C_{11g}\mid1,1,g\rangle + C_{10e}\mid1,0,e\rangle + C_{01e}\mid0,1,e\rangle,
\end{aligned}
\end{equation}

Here, $C_{m,n,g(e)}$ denotes the probability amplitude, and $\mid C_{m,n,g(e)}\mid^2$ representing the probability that the system is in the $\begin{vmatrix}m,n,g(e)\end{vmatrix}$, where $m,n,g(e)$ denotes the number of photons in the cavity and the state of the atom, respectively. To express the photon number distribution in the CW mode in a more straightforward way, $P(m)=\mid C_{m,n,g(e)}\mid^2$ is used to denote the probability that the CW mode has photons. Substituting Eq. (\ref{psi_state}) into the Schrödinger equation $i\partial|\psi(t)\rangle/\partial t=H_{\mathrm{~eff}}|\psi(t)\rangle$ yields the following linear equations:

\begin{equation}
\begin{aligned}
& i\dot{C}_{00g} = \Omega C_{10g}, \\
& i\dot{C}_{10g} = \Omega C_{00g} + \Delta_1 C_{10g} + g C_{00e}, \\
& i\dot{C}_{01g} = \Delta_1 C_{01g} + g C_{00e}, \\
& i\dot{C}_{00e} = g C_{10g} + g C_{01g} + \Delta_e C_{00e}, \\
& i\dot{C}_{20g} = \sqrt{2}\Omega C_{10g} + 2(\Delta_1+\chi) C_{20g} + \sqrt{2}g C_{10e}, \\
& i\dot{C}_{02g} = 2(\Delta_1+\chi) C_{02g} + \sqrt{2}g C_{01e}, \\
& i\dot{C}_{11g} = \Omega C_{01g} + 2\Delta_1 C_{11g} + g C_{10e} + g C_{01e}, \\
& i\dot{C}_{10e} = \Omega C_{00e} + \sqrt{2}g C_{20g} + g C_{11g} + \Delta_2 C_{10e}, \\
& i\dot{C}_{01e} = \sqrt{2}g C_{02g} + g C_{11g} + \Delta_2 C_{01e},
\end{aligned}
\end{equation}

where $\Delta_1=\Delta-i\kappa/2$, $\Delta_e=\Delta-i\gamma/2$, and $\Delta_2=2\Delta-i(\gamma+\kappa)/2$. Under the weak driving conditions 

$\begin{aligned}[t]
&\{\mid C_{20g}\mid,\mid C_{02g}\mid,\mid C_{11g}\mid,\mid C_{10e}\mid,\mid C_{01e}\mid\} \\
&\qquad \ll \{\mid C_{10g}\mid,\mid C_{01g}\mid,\mid C_{00e}\mid\} \ll \mid C_{00g}\mid,
\end{aligned}$

the system remains in the ground state with high probability, i.e., $i\dot{C}_{00g}=0$. To further simplify calculations, let $\gamma=\kappa$, i.e., $\tilde{\Delta}=\Delta-i\kappa/2=\Delta_e=\Delta_1$, $\Delta_2=2\tilde{\Delta}$, then one can easily obtain the following equations:

\begin{equation}
\begin{aligned}
& 0 = \Omega + \tilde{\Delta} C_{10g} + g C_{00e}, \\
& 0 = \tilde{\Delta} C_{01g} + g C_{00e}, \\
& 0 = g C_{10g} + g C_{01g} + \tilde{\Delta} C_{00e}, \\
& 0 = \sqrt{2}\Omega C_{10g} + 2(\tilde{\Delta}+\chi) C_{20g} + \sqrt{2}g C_{10e}, \\
& 0 = 2(\tilde{\Delta}+\chi) C_{02g} + \sqrt{2}g C_{01e}, \\
& 0 = \Omega C_{01g} + 2\tilde{\Delta} C_{11g} + g C_{10e} + g C_{01e}, \\
& 0 = \Omega C_{00e} + \sqrt{2}g C_{20g} + g C_{11g} + \tilde{\Delta} C_{10e}, \\
& 0 = \sqrt{2}g C_{02g} + g C_{11g} + \tilde{\Delta} C_{01e}.
\end{aligned}
\end{equation}

Solving the above system of equations gives the following steady-state solution for the probability amplitude.

\begin{equation}
C_{00e}=\frac{g\Omega}{\tilde{\Delta}^2-2g^2},
\end{equation}

\begin{equation}
C_{01g}=-\frac{g^2\Omega}{\tilde{\Delta}(\tilde{\Delta}^2-2g^2)},
\end{equation}

\begin{equation}
C_{10g}=\frac{\Omega(\tilde{\Delta}^2-g^2)}{\tilde{\Delta}(2g^2-\tilde{\Delta}^2)}=-\frac{\Omega(\tilde{\Delta}^2-g^2)}{\tilde{\Delta}(\tilde{\Delta}^2-2g^2)},
\end{equation}

\begin{equation}
C_{20g}=\frac{\sqrt{2}\Omega^2}{2}\cdot\frac{D}{F},
\end{equation}

\begin{equation}
C_{11g}=\frac{g^2\Omega^2}{2\tilde{\Delta}(\tilde{\Delta}^2-2g^2)}\cdot\frac{4\tilde{\Delta}^2-3\tilde{\Delta}\chi-2g^2}{2\tilde{\Delta}^3-2\tilde{\Delta}g^2-2\tilde{\Delta}^2\chi+g^2\chi},
\end{equation}

\begin{equation}
C_{10e}=-\frac{1}{B}\left(gC_{11g}+\Omega\left(C_{00e}-\frac{2g}{A}C_{10g}\right)\right),
\end{equation}

\begin{equation}
C_{01e}=-\frac{g}{B}C_{11g},
\end{equation}

\begin{equation}
C_{02g}=-\frac{\sqrt{2}g}{A}C_{01e},
\end{equation}

where $D=Q(\tilde{\Delta}^2-2g^2)-2\tilde{\Delta}(\tilde{\Delta}Q/g^2+(\tilde{\Delta}-\chi))(2\tilde{\Delta}^2-g^2)$, $F=\tilde{\Delta}Q(\tilde{\Delta}^2-2g^2)(2(\tilde{\Delta}-\chi)+\tilde{\Delta}Q/g^2)$, $A=2(\tilde{\Delta}-\chi)$, $B=2\tilde{\Delta}-2g^2/A$, $Q=2[g^2-2\tilde{\Delta}(\tilde{\Delta}-\chi)].$. 

\begin{figure}[b]
    \centering
    \includegraphics[width=0.8\columnwidth]{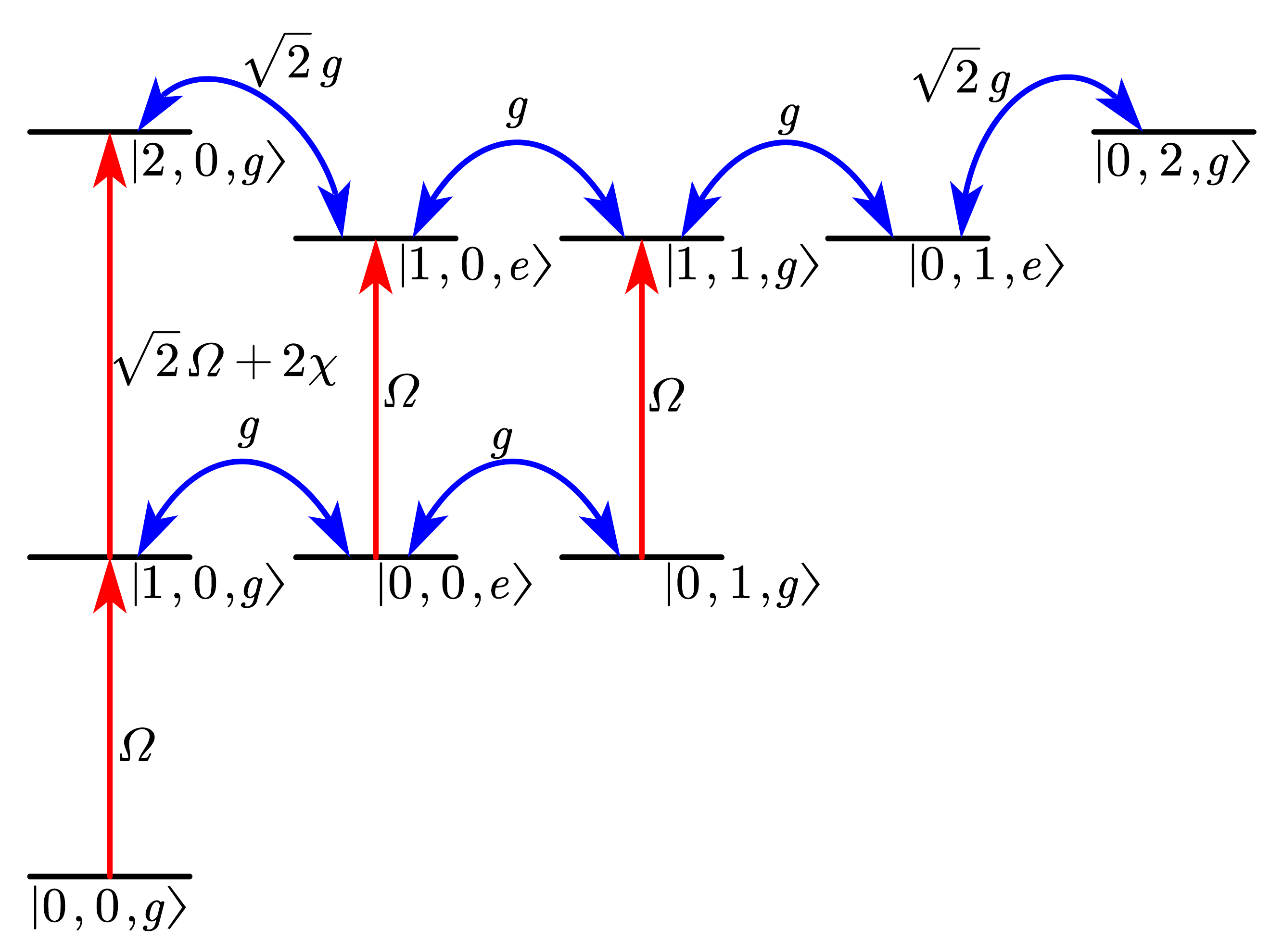}
    \caption{The energy level diagram of the driven CW propagating cavity mode in the WGM within the few-photon subspace. Destructive interference between these paths prevents two photons from occupying the same energy level, leading to UPB.}
    \label{fig:energy-level-diagram}
\end{figure}

The second-order correlation function of the system can be analytically calculated as:

\begin{equation}
\begin{aligned}
g^{(2)}(0) &= \frac{\langle a_1^\dagger a_1^\dagger a_1 a_1\rangle}{\langle a_1^\dagger a_1\rangle^2} = \frac{2\mid C_{20g}\mid^2}{\left(\mid C_{10g}\mid^2 + \mid C_{10e}\mid^2 + \mid C_{11g}\mid^2\right)^2} \\
&\approx \frac{2\mid C_{20g}\mid^2}{\mid C_{10g}\mid^4}.
\end{aligned}
\end{equation}

The perfect PB condition can be obtained by calculating \( |C_{20g}|^2 = 0 \). Since the system can reach the \(|2, 0, g\rangle\) state via three pathways shown as Fig. \ref{fig:energy-level-diagram}, the destructive quantum interference between these pathways suppresses the population of the \(|2, 0, g\rangle\) state, leading to UPB. The three pathways are:

\begin{align*}
\text{a)} &\quad |1, 0, g\rangle \xrightarrow{\sqrt{2}\Omega + 2\chi} |2, 0, g\rangle \\
\text{b)} &\quad |1, 0, g\rangle \xrightarrow{g} |0, 0, e\rangle \xrightarrow{\Omega} |1, 0, e\rangle \xrightarrow{\sqrt{2}g} |2, 0, g\rangle \\
\text{c)} &\quad \begin{aligned}[t]
|1, 0, g\rangle \xrightarrow{g} |0, 0, e\rangle \xrightarrow{g} |0, 1, g\rangle \xrightarrow{\Omega} |1, 1, g\rangle & \\
\xrightarrow{g} |1, 0, e\rangle \xrightarrow{\sqrt{2}g} |2, 0, g\rangle &
\end{aligned}
\end{align*}

\begin{figure}[h]
    \centering
    \includegraphics[width=0.84\columnwidth]{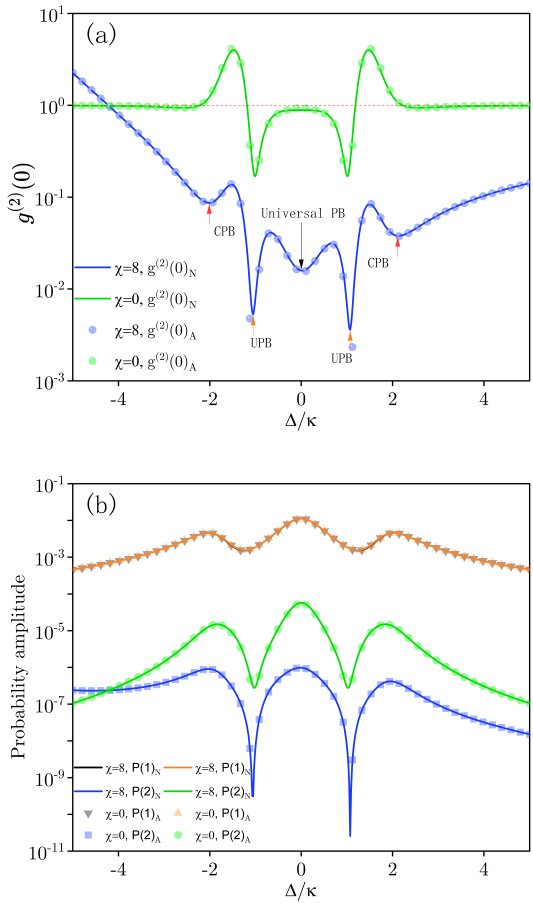}
    \caption{Comparison of numerical and analytical results of the second-order correlation function and the probability amplitudes of photon excitations with and without Kerr nonlinearity (The subscripts “N” and “A” denote “Numerical” and “Analytical”, respectively). (a) Steady-state second-order correlation function $g^{(2)}(0)$ vs. detuning $\Delta/\kappa$. (b) Variations in $P(1)$ and $P(2)$ with detuning. The location of CPB coincides with the maximum value of $P(1)$, while the location of UPB corresponds to the minimum value of $P(2)$. Other parameters: $g/\kappa=1.33$, $\Omega/\kappa=0.1$, $\chi/\kappa=8$, $\gamma=\kappa$. }
    \label{fig:numerical}
\end{figure}

To demonstrate the photon antibunching, the second-order correlation function is plotted as a function of detuning with and without Kerr nonlinearity using numerical and analytical methods, respectively, in Fig. \ref{fig:numerical}(a). As one can see, the analytical results are in good agreement with the numerical simulations. Importantly, the values of second-order correlation functions with strong Kerr nonlinearity are significantly smaller than those without Kerr nonlinearity, implying much stronger photon antibunching. In order to further validate these results, the probability amplitudes of single and two-photon transitions are plotted in Fig. \ref{fig:numerical}(b). As can be easily seen, the three peaks of single-photon resonance transitions match exactly with the three valleys (one of which is labeled as “Universal PB” due to the facts presented in the next section) of the second-order correlation functions, which conforms well with CPB characteristics. Furthermore, two pronounced deep valleys in blue lines appear near the detunings $\Delta=\pm1.05$ (Fig. \ref{fig:numerical}(a)), which can be explained as UPB, since the two-photon population probability significantly decreases at the same detunings (the blue valleys in Fig. \ref{fig:numerical}(b)). Apparently, the optimal detunings between the driving field and cavity mode for UPB are almost not affected by the increase of the Kerr nonlinearity. These results fully confirm that introducing strong Kerr nonlinearity effectively enhances the system's PB characteristics by improving the single-photon purity while preserving the single-photon brightness.

\section{UNIVERSAL PHOTON BLOCKADE}
\label{sec:UNIVERSAL PHOTON BLOCKADE}

\begin{figure*}[t]
        \centering
        \includegraphics[width=0.75\linewidth]{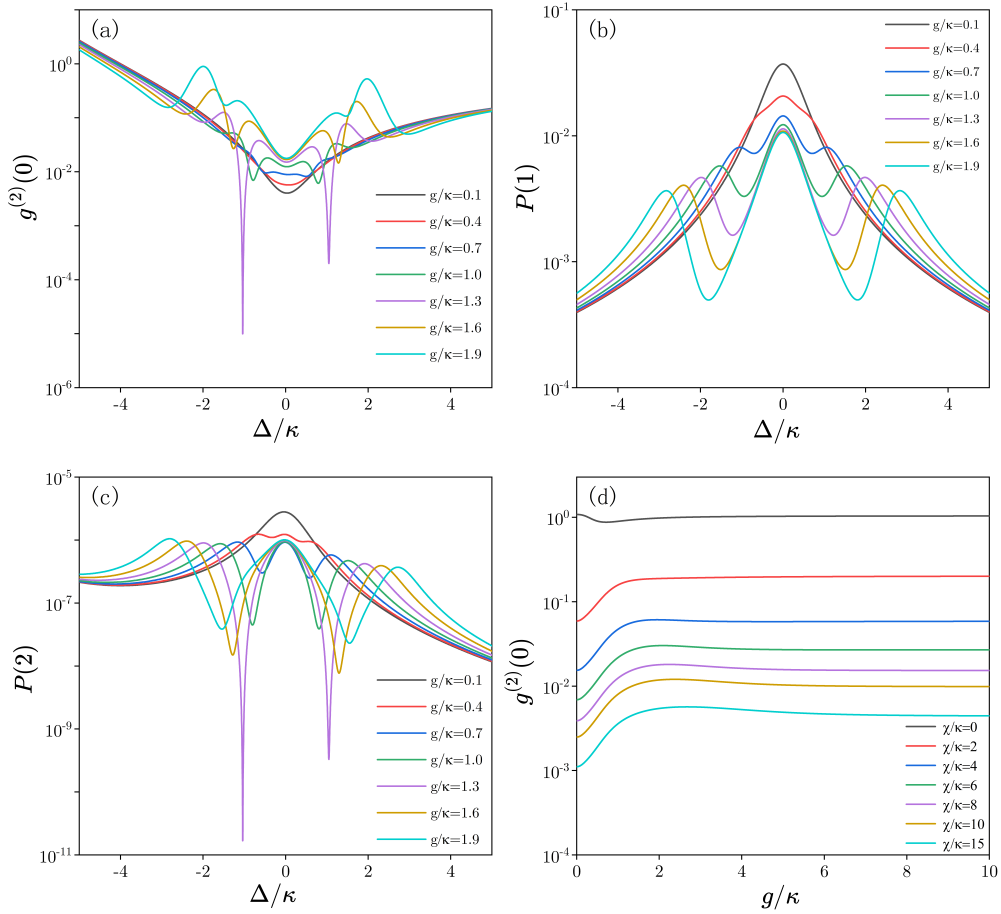}
        \caption{Schematic of the synergistic dual-mechanism resonance position. (a), (b), and (c) show the second-order correlation function of the CW mode, the single-photon probability distribution, and the two-photon probability distribution as a function of detuning and the coupling strength, respectively. The probability distribution of single photons retains its maximum at the resonance position, Fig. 6(b). As the coupling strength decreases, the optimal detuning for UPB approaches the resonance position more closely, Fig. 6(c). The resonance position is simultaneously influenced by the anharmonicity of energy levels and UPB. Fig. 6(d) shows the second-order correlation function at the resonance position as a function of coupling strength and the Kerr nonlinearity. As Kerr nonlinearity becomes stronger, the blockade effect at the resonance position intensifies. For non-zero Kerr nonlinearity, the second-order correlation function first increases, then decreases, and finally stabilizes as the coupling strength grows. Other parameters are the same as in Fig. \ref{fig:numerical}.}

        \label{fig:universal-PB-mechanism}
\end{figure*}

In this section, we demonstrate the antibunching properties of the light field at the resonant position and reveal the underlying physical mechanism. The optimal detuning condition Eq. (\ref{eq:optimal_detuning}) indicates that the resonant position corresponds to one of the optimal parameters for CPB, where the anharmonicity of the energy levels increases with stronger coupling (Fig. \ref{fig:excitation}(a)). However, the second-order correlation function at the resonant position exhibits an anomalous behavior as the coupling strength increases: it first increases, then decreases, and eventually stabilizes, as shown in Fig. \ref{fig:universal-PB-mechanism}(d). This behavior contradicts the conventional understanding that enhanced energy-level anharmonicity leads to stronger antibunching.

To explain this anomalous phenomenon, we examine the blockade characteristics of the system near the optimal coupling strength ($g/\gamma=1.33$) for UPB. As shown in Figs. \ref{fig:universal-PB-mechanism}(a)-\ref{fig:universal-PB-mechanism}(c), the resonant position is very close to the optimal detuning for UPB, leading to a cooperative effect of CPB and UPB at this resonant position, resulting in effective antibunching of the system. As the coupling strength further increases ($g/\gamma>2.2$), although the influence of UPB at the resonant position weakens, CPB still persists and further enhances the blockade characteristics of the system with increasing coupling strength, resulting in a slight decrease in the second-order correlation function.

Within the strong coupling regime ($g\gg\kappa$), the system's blockade characteristics are almost entirely governed by Kerr nonlinearity. However, the second-order correlation function no longer decreases with the coupling strength but converges to a stable value instead, Fig. \ref{fig:universal-PB-mechanism}(d). This behavior cannot be solely explained by CPB, as it indicates that the light field's antibunching reaches saturation even as the anharmonicity of energy levels keeps increasing, Fig. \ref{fig:excitation}(a). The physical origin of this saturation behavior is related to the inherent saturation characteristics of a single two-level atom's response to the light field. As pointed out by Chang et al. \cite{43}, two-level atoms exhibit highly nonlinear optical responses; thus, such atoms cannot simultaneously absorb or emit multiple photons since the absorption of a single photon fundamentally alters the atom's response to subsequent photons. This “single-photon saturation” mechanism implies the upper limit of the atom's response capacity: once excited, the atom cannot interact with another photon again within its finite excited-state lifetime. Under strong coupling conditions, even when the average photon number in the cavity is much smaller than 1, the two-level atom reaches response saturation on the timescale of single-photon interactions, rendering it unable to further regulate the quantum statistical properties of the light field via nonlinear mechanisms.

Srinivasan and Painter observed a similar saturation behavior in their quantum dot-cavity experiments: when the driving intensity exceeds the saturation threshold, the system's photon statistics stabilize and exhibit no significant variation with the coupling strength \cite{23,42}. Our analytical results (Eqs. (\ref{eq:C10g})(\ref{eq:C11g})(\ref{eq:C20g})) further confirm these findings: under the strong-coupling limit at the resonance position, the two-photon occupation probability is primarily determined by the driving intensity $\Omega$, the cavity dissipation $\kappa$, and Kerr nonlinearity $\chi$, independent of the coupling strength $g$, quantitatively revealing the dominant role of the saturation mechanism.

\begin{equation}
\mid C_{01g}\mid^2 \approx \mid C_{10g}\mid^2 \approx \frac{\Omega^2}{\kappa^2},
\label{eq:C10g}
\end{equation}

\begin{equation}
\mid C_{11g}\mid^2 \approx \frac{\Omega^4}{\kappa^2(\chi^2 + \kappa^2)}
\label{eq:C11g}
\end{equation}

\begin{equation}
\mid C_{20g}\mid^2 = \mid C_{02g}\mid^2 \approx \frac{\Omega^4}{2\kappa^2(\chi^2 + \kappa^2)}.
\label{eq:C20g}
\end{equation}

In summary, the anomalous antibunching phenomenon at the resonance position cannot be explained by a single physical mechanism. In the weakly coupled region, the joint effects of the quantum destructive interference between different paths and non-harmonic energy levels cause an initial increase and then a decrease of the second-order correlation function. In the strongly coupled region, the saturation characteristics of atoms dominate, stabilizing the blockade effect, and indicating that the two mechanisms act synergistically rather than independently, and that saturation properties provide additional system robustness under strong coupling. 

\begin{figure}[t]
        \centering
        \includegraphics[width=1\linewidth]{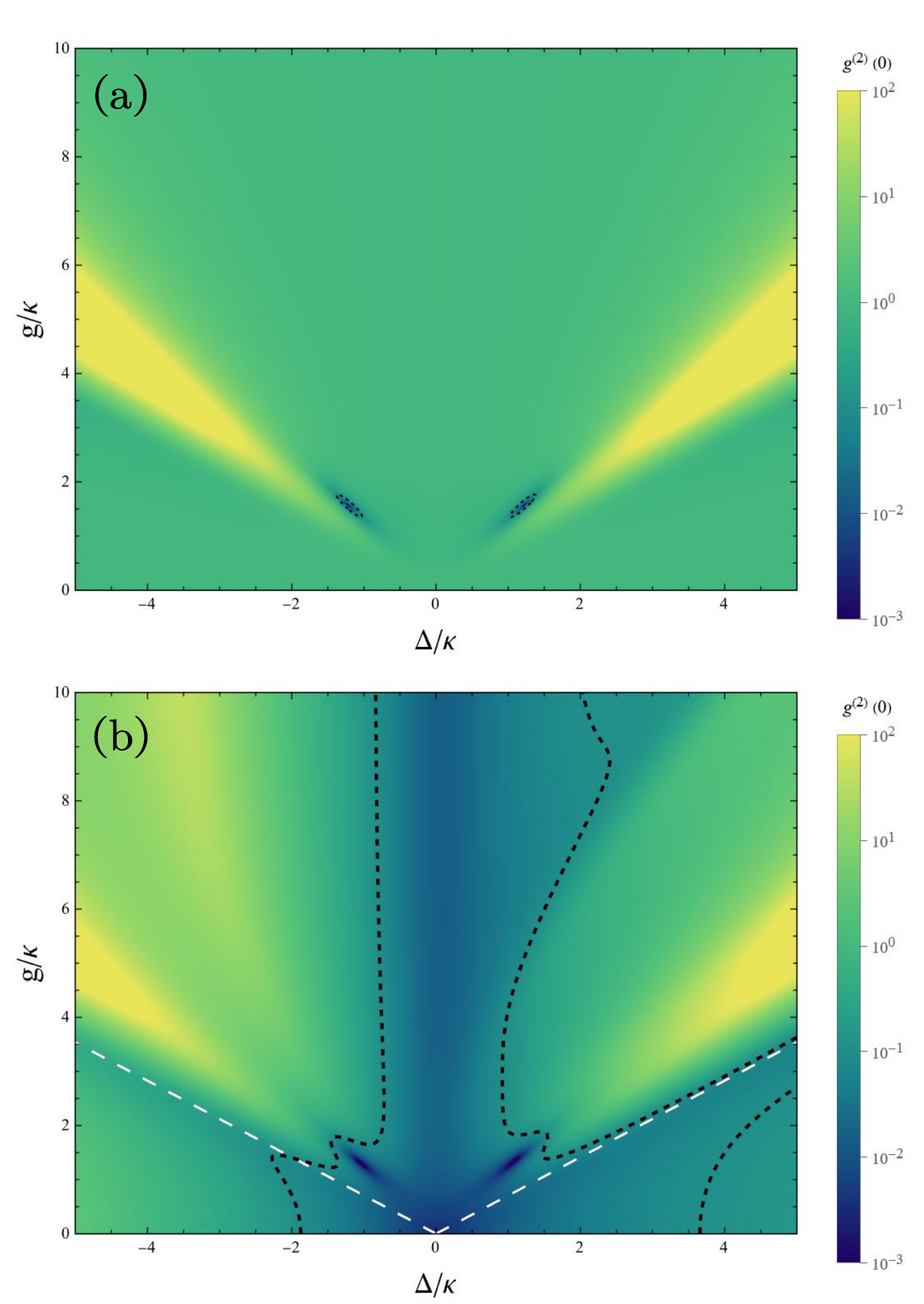}
        \caption{Schematic of universal PB. Density plots of the second-order correlation function as a function of detuning $\Delta/\kappa$ and the coupling strength $g/\kappa$. (a) $\chi/\kappa=0$; (b) $\chi/\kappa=8$. Black dashed lines indicate contour lines for $g^{(2)}(0)=0.1$. White dashed lines denote the optimal parameter $\Delta=\pm\sqrt{2}g$ for CPB. At $\Delta/\kappa=0$, the system consistently exhibits antibunching, demonstrating universal PB properties. Other parameters are identical to those in Fig. \ref{fig:numerical}. }

        \label{fig:universal-PB}
\end{figure}

To more clearly illustrate the PB phenomenon in the system, we plot the second-order correlation function as a function of detuning and coupling strength, as shown in Fig. \ref{fig:universal-PB}. UPB occurs only in the weakly coupled regime, with its optimal parameters symmetrically distributed around the resonance position. Compared with the case without Kerr nonlinearity, the system exhibits effectively enhanced blockade characteristics. In particular, at the resonance position, the PB effect is independent of the coupling strength, manifesting the features of universal PB \cite{37}. Meanwhile, the resonance position consistently maintains high single-photon brightness, as shown in Fig. \ref{fig:universal-PB-mechanism}(b), and enables the preparation of high-purity single photons even under weak coupling conditions, which is highly favorable for experimental implementation.

\section{Universal Photon Blockade with Multi-Parameter Robustness}
\label{sec:Universal Photon Blockade with Multi-Parameter Robustness}

The synergistic advantages of dual mechanisms in enhancing both the purity and brightness of single photons were demonstrated, and the physical mechanism by which resonance positioning achieves universal PB was analyzed. To illustrate the general character of this universal PB, the effects of atomic spontaneous emission rate, drive intensity, and cavity mode interactions on universal PB were investigated.

\subsection{Robustness of universal photon blockade to spontaneous emission rates and drive intensities of atoms}
\label{Robustness of universal photon blockade to spontaneous emission rates and drive intensities of atoms}

\begin{figure}[h]
        \centering
        \includegraphics[width=1\linewidth]{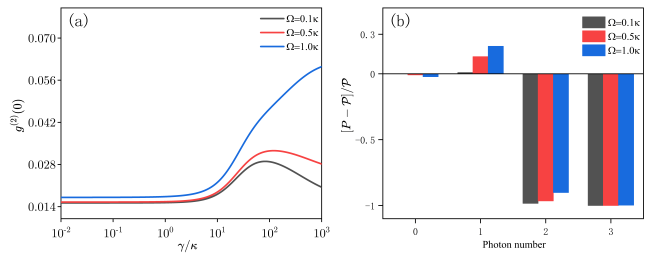}
        \caption{(a) Variation in the second-order correlation function $g^{(2)}(0)$ at the resonance position with respect to the spontaneous emission $\gamma/\kappa$ of the atom. (b) Comparison of the photon number probability distribution $P\begin{pmatrix}m\end{pmatrix}$ in the CW mode with the corresponding Poisson distribution $\mathcal{P}(m)$. A ratio greater than 1 indicates enhancement, while less than 1 indicates suppression. Other parameters are the same as in Fig. \ref{fig:numerical}.}

        \label{fig:Robustness1}
\end{figure}

When the cavity and the atom are in resonance, the Purcell effect significantly enhances the spontaneous emission rate of the atom\cite{49,50}. Therefore, investigating the influence of the atom's spontaneous emission rate on universal PB is essential. Fig. \ref{fig:Robustness1}(a) shows the steady-state second-order correlation function as a function of the atom's spontaneous emission rate at the resonance position, demonstrating that the atom's spontaneous emission rate exhibits good robustness to the antibunching effect of multiple nonlinearities of the optical field. This behavior allows the extension of the developed model beyond real atoms, enabling the study of the universal PB in various physical systems, e.g., solid-state quantum systems and microwave systems. Furthermore, the saturated response characteristics of a single two-level atom to the optical field under strong coupling enable the developed system to tolerate stronger driving intensities, even when they have the same order of magnitude as the atom's dissipation ($\Omega\sim\kappa$). Fig. \ref{fig:Robustness1}(b) illustrates the statistical properties of the cavity field determined by calculating the relative deviation $[P(m)-\mathcal{P}(m)]/\mathcal{P}(m)$ between the cavity photon number distribution $P(m)$ and a Poisson distribution $\mathcal{P}(m)\equiv\langle m\rangle^me^{-\langle m\rangle}/m!$. Under strong coupling, without exceeding the saturation threshold, the probability $P(1)$ of detecting a single photon significantly increases with the drive intensity, while the probability of detecting $m\geq2$ photons is strongly suppressed.

\subsection{Robustness of universal photon blockade to defect-induced mode coupling}
\label{Robustness of universal photon blockade to defect-induced mode coupling}

\begin{figure}[h]
    \centering
    \includegraphics[width=1.0\columnwidth]{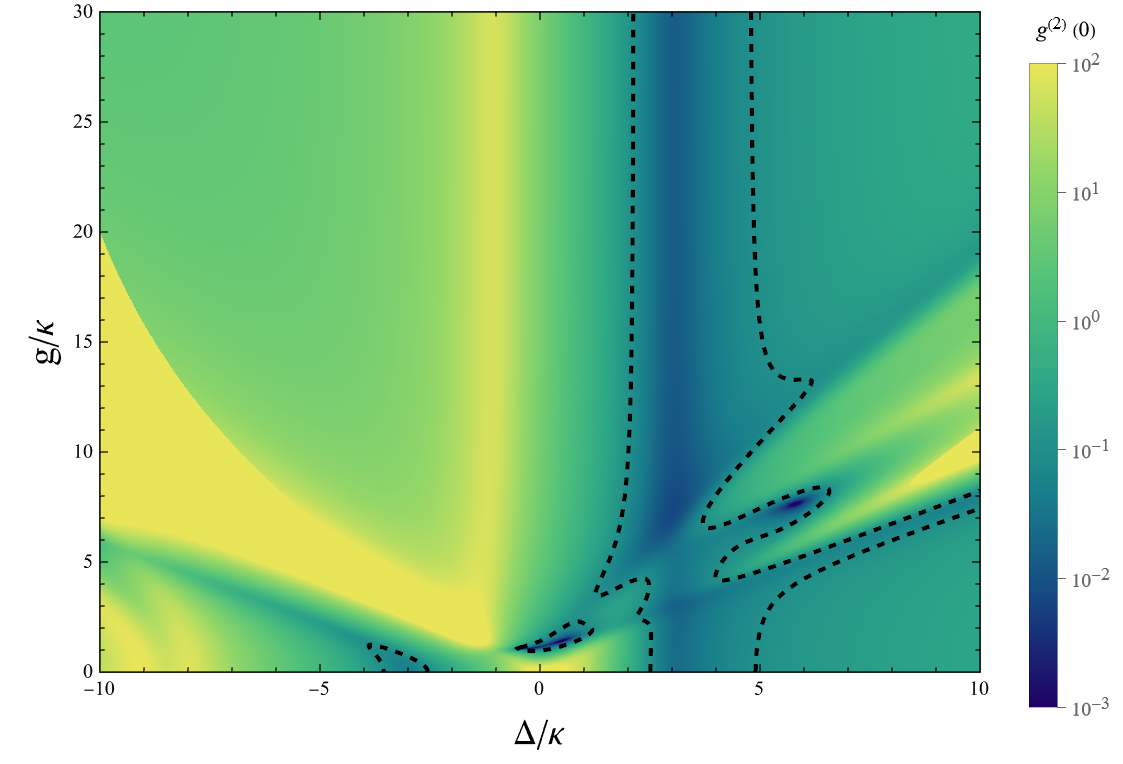}
    \caption{The density plot of the second-order correlation function as a function of detuning and the cavity-atom coupling strength g with coupling between cavity modes ($J=3\kappa$). The black dashed curve represents the second-order correlation function $g^{(2)}(0)=0.1$, where the detuning position corresponding to universal PB is $\Delta=J$. Other
parameters are identical to those in Fig. \ref{fig:numerical}.}
    \label{fig:Robustness2}
\end{figure}

In a real WGM cavity, intrinsic material defects, density variations, or surface roughness can backscatter light in the reverse propagation direction, introducing additional optical loss and inter-mode coupling\cite{51,52}. Here, $\hat{H}_{int}=J(\hat{a}_1^\dagger\hat{a}_2+\hat{a}_2^\dagger\hat{a}_1)$ is used to describe the strength of this inter-mode coupling; the system's Hamiltonian can be rewritten as $\hat{H}_{tot}=\hat{H}+\hat{H}_{int}$. The single-excitation dressed state energy levels can be obtained by solving the corresponding eigenvalue equations: $E_+=\omega+(J+\sqrt{J^2+8g^2})/2$, $E_0=\omega-J$, $E_-=\omega+(J-\sqrt{J^2+8g^2})/2$. Without mode coupling ($J=0$), the above eigenenergies reduce to those of the standard JC model. For finite $\text{J}$, the resonance energy levels experience a shift proportional to $\text{J}$, as shown above. Due to Kerr nonlinearity, the anharmonic nature of the system's energy levels persists. In this scenario, the detuning position where universal PB occurs simply shifts to the $\Delta=J$ position, Fig. \ref{fig:Robustness2}.

\section{CONCLUSIONS}
\label{CONCLUSIONS}

In this paper, we achieved universal PB in a dual-mode JC model with strong Kerr nonlinearity. The results were validated by numerically solving the master equation in the steady-state limit and analytically calculating the second-order correlation function using the probability amplitude method. The analysis of the effects of Kerr nonlinearity on conventional and UPB revealed two mechanisms that synergistically enabled multi-parameter robust universal PB, particularly at resonance positions.

It was demonstrated that (i) CPB can also be achieved at resonance positions in a two-mode JC model with Kerr nonlinearity, compared to a model without Kerr nonlinearity; (ii) UPB primarily occurs under weak coupling conditions ($g/\kappa<2.2$), where Kerr nonlinearity enhances the system's blockade strength by suppressing the probability distribution of two-photon events by increasing the anharmonicity of energy levels. Kerr nonlinearity scarcely alters the optimal parameters for UPB while broadening its parameter range; (iii) under weak coupling conditions, the resonance position lies very close to the optimal detuning for UPB. This proximity causes the synergy of two mechanisms, enabling the system to achieve strong antibunching within the weak coupling regime while maintaining high brightness. As the coupling strength further increases, the PB characteristics become primarily governed by Kerr nonlinearity. At this stage, the blockade does not intensify with the coupling strength-induced increase in the anharmonicity of energy levels. This anomalous behavior originates from the response of the two-level atom to the saturation of the light field. The invariability of the resonance position with the atom-cavity coupling aligns with the defining features of universal PB; (iv) Although the Purcell effect enhances the atom's spontaneous emission rate when the cavity and atom resonate, this universal PB persists. Furthermore, it is also robust against defect-induced inter-mode coupling, although the optimal detuning position for universal PB shifts with the coupling strength between modes ($\Delta=J$). 

In summary, introducing Kerr nonlinearity into the two-mode JC model effectively enhances the purity of single photon while preserving its brightness. This work not only extends the theoretical framework of universal PB but also demonstrates its robust and wide applicability under diverse parameter conditions, further validating its potential for developing high-performance, integrated single-photon quantum optical devices.

\section{ACKNOWLEDGMENTS}
\label{ACKNOWLEDGMENTS}

This work is supported by the National Natural Science Foundation of China (Grant No. 12247101), the Fundamental Research Funds for the Central Universities (Grant No. lzujbky-2025-jdzx07), the Natural Science Foundation of Gansu Province (No.25JRRA799), and the ‘111 Center’ under Grant No. B20063. We also thank Dr. An Junhong from Lanzhou University for his valuable comments and correction on this work.

\section{DATA AVAILABILITY}
\label{DATA AVAILABILITY}

The data that support the ﬁndings of this article are not publicly available. The data are available from the authors upon reasonable request.

\clearpage
\newpage
\bibliography{References}

\end{document}